\input ./style/arxiv-general.cfg
\documentclass[aoas,MSNbibl,nameyear,dvips]{arximspdf}
\makeatletter
   \@ifpackageloaded{graphicx}{}{\usepackage{graphicx}}
\makeatother
\usepackage{textcomp}

\doi{10.1214/15-AOAS870}
\volume{9}
\issue{4}
\pubyear{2015}
\firstpage{1973}
\lastpage{1996}
\docsubty{FLA}

\makeatletter
\newcommand{\colboxcancer}{orange\ }
\newcommand{\colboxstroma}{green\ }
\newcommand{\colboxadipose}{cyan\ }
\newcommand{\colboxofftissue}{pink\ }
\newcommand{\mData}{\mathbb{X}}
\newcommand{\vDataRow}[1]{\mathbf{x}_{#1 \bullet}}
\newcommand{\vDataCol}[1]{\mathbf{x}_{\bullet #1}}
\newcommand{\vTemplate}[1]{\mathbf{t}_{#1}}
\newcommand{\vCentroid}{\mathbf{c}}
\newcommand{\vDistanceNeighbourhood}[1]{\mathbf{d}_{#1}}
\newcommand{\vDIPS}{\mathfrak{d}}
\newcommand{\sData}[2]{x_{#1#2}}
\newcommand{\sSmoothProp}[2]{T_{#1#2}}
\newcommand{\sCutoff}{a}
\newcommand{\subsetCol}{\mathcal{S}}
\newcommand{\argmin}[1]{\mathop{\operatorname{arg}\operatorname{min}}\limits_{#1}}
\makeatother

\begin{document}
\begin{frontmatter}

\title{Feature extraction for proteomics imaging mass spectrometry data}
\runtitle{Feature extraction for IMS}

\begin{aug}
\author[A]{\fnms{Lyron J.}~\snm{Winderbaum}\corref{}\ead[label=e1]{lyron.winderbaum@student.adelaide.edu.au}},
\author[A]{\fnms{Inge}~\snm{Koch}\ead[label=e2]{inge.koch@adelaide.edu.au}},
\author[A]{\fnms{Ove J.~R.}~\snm{Gustafsson}\ead[label=e3]{ove.gustafsson@adelaide.edu.au}},
\author[A]{\fnms{Stephan}~\snm{Meding}\ead[label=e4]{stephan.meding@adelaide.edu.au}}
\and
\author[A]{\fnms{Peter}~\snm{Hoffmann}\thanksref{T1}\ead[label=e5]{peter.hoffmann@adelaide.edu.au}}
\runauthor{Winderbaum et al.}
\affiliation{The University of Adelaide}
\address[A]{School of Mathematical Sciences\\
The University of Adelaide\\
Adelaide SA 5005\\
Australia\\
\printead{e1}\\
\phantom{E-mail: }\printead*{e2}}
\end{aug}
\thankstext{T1}{Supported in part by the Australian Research Council
(ARC LP110100693), Bioplatforms Australia and the
Government of South Australia.}

%
\received{\smonth{6} \syear{2014}}
%
\revised{\smonth{8} \syear{2015}}

%
\begin{abstract}
Imaging mass spectrometry (IMS) has transformed
proteomics by providing an avenue for collecting
spatially distributed molecular data.
Mass spectrometry data acquired with matrix assisted laser
desorption ionization (MALDI) IMS consist of tens of
thousands of spectra, measured at regular grid points
across the surface of a tissue section.
Unlike the more standard liquid chromatography mass
spectrometry, MALDI-IMS preserves the spatial
information inherent in the tissue.

Motivated by the need to differentiate cell
populations and tissue types in MALDI-IMS data
accurately and efficiently, we propose an integrated
cluster and feature extraction approach for such data.
We work with the derived binary data representing
presence/absence of ions, as this is the essential
information in the data.
Our approach takes advantage of the spatial structure
of the data in a noise removal and initial dimension
reduction step and applies $k$-means clustering with
the cosine distance to the high-dimensional binary
data.
The combined smoothing-clustering yields spatially
localized clusters that clearly show the correspondence
with cancer and various noncancerous tissue types.

Feature extraction of the high-dimensional binary data
is accomplished with our difference in proportions of
occurrence (DIPPS) approach which ranks the variables
and selects a set of variables in a data-driven manner.
We summarize the best variables in a single image that
has a natural interpretation.
Application of our method to data from patients with
ovarian cancer shows good separation of tissue types
and close agreement of our results with tissue types
identified by pathologists.
\end{abstract}

%
\end{frontmatter}

\section{Introduction}
\label{intro}

Mass spectrometry (MS) has become a versatile and
powerful tool in proteomics for the analysis of
complex biological systems, including the
identification and quantification of proteins
and peptides [\citet{Ong2005}].
Many different technologies have been developed
under the collective field of proteomics mass
spectrometry [\citet{Aebersold2003}].
The focus in this paper is the more recent
development [see \citet{Groseclose2007,Aoki2007}]
of matrix assisted laser desorption ionization
(MALDI) imaging mass spectrometry (IMS), also
known as MALDI imaging, and, in particular, an
analysis of MALDI-IMS data acquired from
tissue samples of patients with ovarian cancer.

Unlike the more common 2D gel electrophoresis (2D-GE)
and liquid chromatography (LC) based techniques in
proteomics, MALDI-IMS preserves the spatial
distribution inherent in the tissue; and the tens of
thousands of spatially distributed spectra acquired
from a single tissue sample in a MALDI-IMS experiment
provide new challenges for statisticians and
bioinformaticians as well as having the potential to
lead to breakthroughs in biological research
[see \citet{Casadonte2011}].
We propose a combined cluster and feature extraction
method for such data which exhibits cancer-specific
variables whose protein associations can be inferred
by parallel LC-MS experiments such as those of
\citet{Meding2012}.

Standard proteomics mass spectrometry methods such as
$2$D-GE and LC-MS have been described in the
literature for some time; see \citet{Wasinger1995}.
We briefly explain LC-MS and important differences with
MALDI-IMS in Section~\ref{proteomics}.
For an overview and review of recent approaches in
LC-MS, see \citet{America2008}.
Statistical challenges of proteomics mass spectrometry
data are outlined in \citet{Wu2003}.
The statistics and bioinformatics literature on the
analysis of $2$D-GE and LC-MS data is growing fast and
covers a range of statistical methods.
Testing and classification of such data are described
in \citet{Morris2005,Morris2012,Yu2006} and references
therein.
Other statistical approaches that have been proposed
and applied to 2D-GE and LC-MS data include peak
identification, alignment and feature selection
[see \citet{Yu2006}], identification of proteins [see
\citet{Yu2006} and \citet{Karpievitch2010}],
wavelet-based methods [see \citet{Morris2006},
\citet{America2008}, \citet{Du2006} and references
therein] and methods from survival analysis for the
detection of differentially expressed proteins [see
\citet{Tekwe2012}].

Contrasting these developments in the analyses of
2D-GE and LC-MS data, the newer MALDI-IMS methods
which have been introduced into routine research
practice have not yet attracted as much attention in
the statistics literature, although MALDI-IMS methods
are covered in proteomics/\break mass spectrometry journals---see
\citet{Alexandrov2011}, Alexandrov et~al. (\citeyear{Alexandrov2010,Alexandrov2013}),
\citet{Norris2007,Jones2012,Gessel2014,Stone2012}
and references therein.
The potential of MALDI-IMS is described in
\citet{Alexandrov2011}: ``\textit{IMS is one of the most
promising innovative measurement techniques in
biochemistry which has proven its potential in
discovery of new drugs and cancer biomarkers$\ldots.$ IMS
was used in numerous studies leading to understanding
chemical composition and biological processes$\ldots.$ As
for many modern biochemical techniques}, \textit{in particular
in proteomics}, \textit{the development of computational methods
for IMS is lagging behind the technological
progress}.''
In addition to presenting our approach and analyses of
MALDI-IMS data, we hope to motivate other statisticians
and bioinformaticians to explore this exciting and
promising new area and to develop novel statistical
methods for the analysis of such data.

%
\begin{figure}[t]

\includegraphics{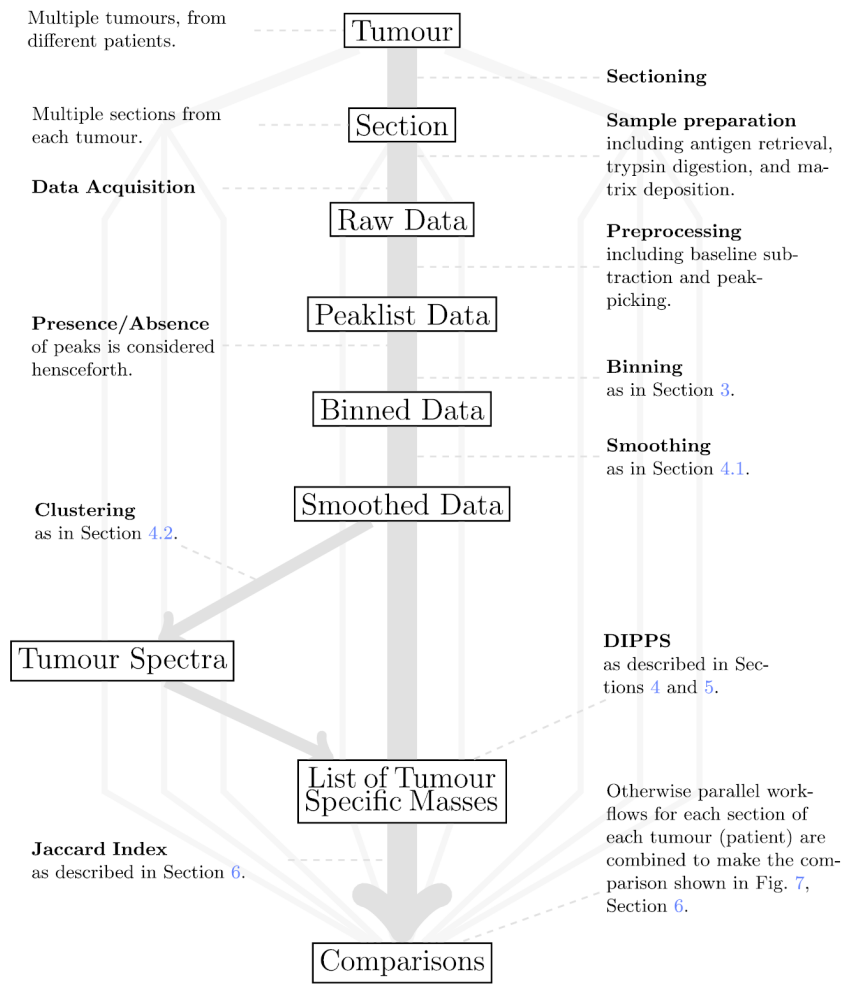}

\caption{Data processing workflow for our proposed
combined clustering-DIPPS approach.
The background arrows (in light grey) represent parallel
workflows for the three sections from each of the three
patients, each leading to a ``list of tumor specific
masses.''
In the final step the lists are compared as described
in Section~\protect\ref{datasetComparison}.}
\label{flowchart}
\end{figure}

Related to our research are the papers by
\citet{Deininger2008,Alexandrov2010} and
\citet{Bonnel2011} who cluster their MALDI-IMS data
using principal component analysis and hierarchical
clustering, or Gaussian mixture models.
Our proposal, outlined in Figure~\ref{flowchart}, differs
from their research in a number of important aspects.
Unlike these authors, we derive suitably binned binary
data, which we describe in Section~\ref{data}, instead of
working with the raw or intensity data.
Following [\citet{Koch2013}, Chapter~6], who demonstrates
the success of using such binary data in finding
biologically meaningful tissue clusters, we apply
$k$-means clustering to the binary data.
Analysis of the binary data has the added advantage of
being computationally more efficient.
Furthermore, parallel computation is possible for all
steps in the approach we propose.
Our approach combines clustering with explicit feature
extraction conceptually similar to the approach of
\citet{Jones2011}, although based on a different
principle---our Difference in Proportions of
Occurrence (DIPPS) statistic which ranks and selects
the ``best'' variables in a data-driven way.
Our use of binary data allows the feature extraction
results to be visualized as a single heat map with a
natural interpretation.
The ability to visualize the selected features as a
single easily interpretable image has not been a part
of the above-mentioned papers and gives a significant
advantage to our approach.

This paper is organized in the following way. We
briefly describe relevant background on proteomics in
Section~\ref{proteomics}, and discuss advantages of
MALDI-IMS for the biological research fields.
We describe MALDI-IMS data and how to derive the binned
binary data in Section~\ref{data}.
Section~\ref{clus} covers the spatial smoothing and
clustering steps of our method.
The DIPPS approach (including a feature extraction
step) is described in Section~\ref{DIPS}.
Finally, in Section~\ref{datasetComparison} we apply our
combined cluster and DIPPS approach in order to compare
several patients by considering several datasets.
The results discussed in Section~\ref{datasetComparison}
allow us to demonstrate how these data could be used to
address biologically relevant questions such as the
identification of potential tissue-specific protein
markers and classification of patients based on their
response to treatment.

\section{Proteomics background}
\label{proteomics}

Ovarian cancers are virtually asymptomatic and, as a
result, the vast majority of cases are detected when the
disease has metastasized.
For these patients, radical surgery and chemotherapy
are often insufficient to address the disease
adequately and many patients relapse.
The combination of late-stage diagnosis and
unsuccessful treatments makes ovarian cancer the most
lethal gynecological cancer, with advanced stage
patients exhibiting a five year survival rate of less
than $30\%$
[\citet{Ricciardelli2009,Jemal2011}].
The keys to addressing ovarian cancer will be as follows:
increasing our understanding of the mechanisms driving
cancer progression, identifying molecular markers which
can predict treatment success and identifying new
treatment targets.
As proteins are key functional components of cells and
tissues, determining protein distributions in cancer
tissue represents a crucial step in addressing these
key aims.

Proteins are synthesized within cells as linear amino
acid sequences and folded into more complex 3D
structures that determine function and intracellular
location.
The complete set of proteins which exist in a given
cell, tissue or biological fluid, under defined
conditions, is termed its proteome [\citet{Wilkins1996}].
Proteomes vary considerably between different cellular
states and understanding these variations allows
insight into the development and progression of cancer.
Proteomics characterizes proteome changes using a
combination of fractionation, identification and
quantitation strategies.
Proteomics will use either a top-down or bottom-up
approach.
Top-down approaches analyze intact proteins, whereas
bottom-up approaches use proteolytic enzymes (e.g.,
trypsin) to digest proteins into peptides prior to
analysis.
The data we present is on tryptic peptides, so our
discussion is in the context of a bottom-up approach.
Proteome fractionation can be achieved using
gel electrophoresis [\citet{Gygi2000}] or liquid
chromatography (LC), with LC being the predominant
fractionation technique
[\citet{Rogowska-Wrzesinska2013}].
LC makes use of columns to affinity-bind molecules to a
stationary phase.
The molecules are subsequently eluted over time with a
changing gradient of mobile phase solvent.
In a bottom-up LC experiment peptides in a hydrophilic
mobile phase are bound to a hydrophobic stationary
phase.
The peptides are eluted using an increase in the
percentage of hydrophobic solvent in the mobile phase.
To characterize the fractioned peptides, the LC eluant
is often directly coupled to an MS instrument (LC-MS).

MS instruments contain an ion source, mass analyzer and
detector.
So-called ``soft'' ionization sources such as
electrospray ionization (ESI) and matrix-assisted laser
desorption/ionization (MALDI) are favored in
proteomics, as they prevent significant molecular
fragmentation during ionization.
LC-MS instruments usually employ ESI to produce gaseous
ions for mass analysis. The mass-to-charge ratio ($m/z$)
of these ions is measured by the mass analyzer and
detector to produce a mass spectrum of measured
intensity as a function of $m/z$.
Peptides analyzed by LC-MS can be fragmented to
produce spectra which correspond to amino acid
composition.
Identification of these spectra is attempted using
search algorithms, such as MASCOT [see
\citet{Koenig2008} and references therein], which match
measured spectra to expected fragmentation spectra.
The combination of LC-MS with bottom-up strategies is
what allows proteomic studies to identify and quantify
thousands of unique proteins in a given biological
sample.

%
\begin{figure}[t]

\includegraphics{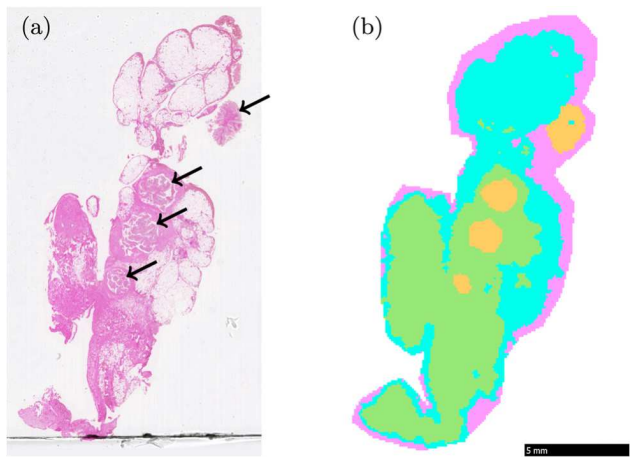}

\caption{Left, \textup{(a)} shows the H$\&$E stained tissue
section, arrows indicate the four cancer nodules.
Right, \textup{(b)} shows pixels plotted in their relative
spatial locations, color identifies the cluster
membership produced by 4-means clustering of the binary
spatially smoothed data.}\label{Cos4clus}
\end{figure}

LC-MS suffers from two
method-specific limitations:
\begin{longlist}[2.]
\item[1.] Tissue samples are homogenized and solubilized,
which removes all spatial information inherent in the
tissue, and

\item[2.] An LC-MS run usually takes more than an hour,
precluding the rapid ($\leq$1~day) analysis of large
sample numbers ($\geq$20).
\end{longlist}

Given that tissues are a mix of different cell
populations and their organization is directly related
to their functions, point 1 above can be crucial.
Typically, tissue structure is visualized by a
pathologist using histological stains such as
haematoxylin and eosin (H$\&$E) followed by light
microscopy, as shown in Figure~\ref{Cos4clus}(a).
Such histological stains allow visualization of the
spatial distribution of cellular morphology and can
provide an understanding of the way in which the
cellular morphology relates to cancer behavior and,
ultimately, the cancer's effect on the patient.
The spatial information is essential in these
histological stains, and it is reasonable to assume it
would be equally crucial in mass spectrometry.
The loss of spatial information that occurs during
sample preparation for LC-MS analysis therefore
motivated the development of direct tissue analysis
using MALDI-IMS
[\citet{Cornett2007,Groseclose2008,Gustafsson2011}].
To prepare a sample for MALDI-IMS, a tissue block is
thinly sectioned (2--10~\textmu m thick slices) and
mounted onto conductive microscopy slides.
For analysis of peptides, the tissue section is coated
with a homogeneous layer of proteolytic enzyme to
digest endogenous proteins.
The digest is then overlaid with a matrix compound
which co-crystallizes with the tissue-derived peptides.
A MALDI source is used to ionize these peptides
directly from the tissue and facilitate the collection
of a mass spectrum from each $(x,y)$ position on a
regular grid across the tissue section.
An acquisition spacing of 20--250~\textmu m is typical
and balances the requirements for high quality MS data,
practical data size and measurement time.
In the datasets presented here, we used a spacing of
100~\textmu m.
The availability of spatial information in MALDI-IMS
offers a unique perspective on tissue analysis, which
is complementary to LC-MS.

In this context, the advantages of MALDI-IMS are as
follows:
\begin{longlist}[3.]
\item[1.] Hundreds of biological molecules can be
measured in a single experiment in an untargeted
manner, as compared to immunohistochemistry.

\item[2.] Tissue sections can be H$\&$E stained
post-acquisition of MALDI-IMS data
[\citet{Deutskens2011}] and spatial changes in protein
abundance measured by MALDI-IMS can be compared to
the histology visible in the H$\&$E stain of the same
tissue (see Figure~\ref{Cos4clus}).
This makes the technique compatible with pathology.

\item[3.] MALDI-IMS provides a capacity for rapid
interrogation of large sample numbers.
This can be achieved using tissue microarrays (TMAs) [see \citet
{Groseclose2008,Steurer2013}];
construction of TMAs involves the extraction of
cylindrical cores from multiple tissue blocks and
arrangement of these cores within a new block.
A one-day sample preparation of sections from such a
block allows an overnight MALDI-IMS experiment to
collect data for $\geq$50 samples.
\end{longlist}

A caveat of MALDI-IMS analyses is that the proteome cannot be fractionated, and the masses measured using
MALDI-IMS are not fragmented to provide sequence
information.
LC-MS based proteomics is required in order to infer
parent proteins, as in Table~\ref{tabmassIDs}.
MALDI-IMS, however, places this information
in a structural context and is therefore a crucial link
between molecular composition and morphology.

%
\begin{table}[b]
\tabcolsep=0pt
\caption{Selection of peptide $m/z$ values, the DIPPS
score of the $m/z$ bins containing them,
and their inferred parent proteins. Peptide sequences
and parent proteins were inferred by mass  matching to
concurrent LC-MS/MS analyses and validated by both
{in situ} MALDI-MS/MS and immunohistochemistry as
shown in Supplement \textup{A} [\citet{IHC}]}\label{tabmassIDs}
\begin{tabular*}{\tablewidth}{@{\extracolsep{\fill}}@{}lccc@{}}
\hline
\textbf{LC-MS/MS} & \textbf{DIPPS} & \textbf{UniProtKB/SwissProt} & \textbf{Protein} \\
\textbf{mass [M${}\bolds{+}{}$H]$^{\bolds{+}}$} & \textbf{statistic} & \textbf{entry name} & \textbf{name}\\
\hline
1628.8015 & 0.662 & ROA1\_HUMAN & Heterogeneous nuclear \\
& & & ribonucleoprotein A1 \\[3pt]
2854.3884 & 0.910 & K1C18\_HUMAN & Keratin 18\\
\hline
\end{tabular*}
\end{table}

MALDI-IMS therefore provides unprecedented capacities
for both proteomics and pathology and, as a consequence,
has important implications for research into human
cancers [\citet{Gorzolka2014}].
For example, ovarian cancers are known to be quite
heterogeneous tissues [\citet{Deininger2008}].
Spatial analysis by MALDI-IMS allows this heterogeneity
to be addressed explicitly [\citet{Gorzolka2014}].
Furthermore, if TMAs are used, hundreds of patients can
potentially be analyzed in a single day.
MALDI-IMS provides the opportunity to (i) screen
preprepared cancer samples rapidly, targeting regions
of interest and, (ii)~employ complementary LC-MS methods
to characterize the tumor proteome of these tissues.

To exploit these capabilities in the future, it will be
crucial to understand and distinguish differences
between patients.
As proof of the concept, in this paper we take data
including multiple tissue types (e.g., tumor, adipose,
connective) from full sections of tissue from each of
three patients, and consider the problem of separating
tumor-specific masses.

\section{The data and binary binning}
\label{data}

In Section~\ref{datasetComparison} we consider a selection
of datasets from three surgically excised ovarian
cancers, each from a different patient; see
\citet{Gustafsson2012thesis}.
An overview for our data analysis workflow is described
in Figure~\ref{flowchart}.
Thin cross-sections of tissue are obtained from each
cancer, and MALDI-IMS data are collected using a Bruker
Ultraflex III and a $m/z$ range of 1000--4500.
Multiple sections are taken from each cancer, and we
refer to the data collected from each section as a
``dataset.''
Initially, we consider a single typical dataset (shown
in Figure~\ref{Cos4clus}) and will refer to it as the
``motivating dataset'' throughout.
A typical dataset might have 5000--100,000 spectra, the
{motivating dataset} consists of
13,916.
The {motivating dataset} will serve to illustrate our
proposed method which we describe in Sections~\ref{clus} and \ref{DIPS}, completing the
workflow illustrated in Figure~\ref{flowchart}.
In Section~\ref{datasetComparison} we consider the remaining
datasets (represented by grey background arrows in
Figure~\ref{flowchart} and including the motivating dataset),
three from each cancer, and compare datasets within and
between patients.
Although we have access to more datasets, we show only
nine, as these suffice for comparisons both within and
between patients while still illustrating our method
concisely.

As discussed in Section~\ref{proteomics}, each dataset
contains mass spectra collected at regularly spaced
points on the surface of a thin tissue section.
Each mass spectrum has annotation meta-data
corresponding to the spatial $(x,y)$ coordinates of its
acquisition.
The mass spectrum itself is a set of ion-intensity,
$m/z$ value pairs, and can be thought of as a discrete
approximation of ion-intensity as a function of $m/z$.
We will not explicitly address this functional aspect.
Each mass spectrum typically contains between $5$ and
$200$ intensity-peaks (local maxima), each
corresponding to the $m/z$ of a biological analyte such
as a peptide or endogenous protein.
A number of preprocessing steps are involved in
extracting the peaks; we use methods
available in proprietary software (flexAnalysis, Bruker
Daltonik, \surl{http://www.bruker.com}): smoothing
(Gaussian kernels), baseline reduction (TopHat), and,
finally, peak-picking (SNAP).
The SNAP algorithm isolates mono-isotopic peaks and
defines significant peaks as those peaks with a
signal-to-noise ratio of two or higher.
Representing the data as peak lists significantly
reduces the amount of data involved, often by as much
as two orders of magnitude, and this can be very
important due to the amount of data involved ($>$10~GB).
We will be concerned only with these extracted peaks
from here on.

Improved peak-picking is of interest for two reasons.
First, peak-picking is currently the most
computationally intensive step in our workflow, taking
longer than the remainder of the data analysis
combined.
Conveniently, the Bruker peak-picking can be run
simultaneously with the data acquisition, which
typically takes 6--14 hours depending on the number
of spectra acquired.
Second, the choice of peak-picking algorithm will affect
all subsequent data analysis and so improvements to
the peak-picking algorithm would be expected to carry
through and produce improved results downstream.
Comparing the Bruker peak-picking that we have used
with other existing methods and making improvements
based on these comparisons is important, but is beyond
the scope of this paper.
All the code used to generate our results from the
peaklist data is included in Supplement~B [\citet{code}], and the
peaklist data itself is in Supplement~C [\citet{peaklistdata}].
Software packages that implement existing peak-peaking
algorithms and other relevant analysis tools include
the R package MALDIquant
(\surl{http://strimmerlab.org/software/maldiquant/}),
the MATLAB bioinformatics toolbox
(\surl{http://au.mathworks.com/help/bioinfo/\\index.html})
and Bioconductor (\surl{http://www.bioconductor.org}).
For other\break mass-spectrometry based
software resources, see \surl{http://www.ms-utils.org}.

We bin the peaks by partitioning the $m/z$ range into
equal size intervals or bins, and identifying each peak
with the bin it belongs too.
Using a data-independent partitioning for the binning
makes comparisons between datasets straightforward.
The disadvantage is that peaks detected at very similar
$m/z$ values will be identified with different bins if
a boundary happens to fall between them.
In order to address this, we suggest running all
analyses in tandem with shifted bin locations and
combining the results in order to capture any analytes
whose $m/z$ is too close to a bin boundary.
Using larger bin sizes also helps limit this effect.
For the sake of brevity, we will omit the tandem
analysis from the results, as it provides only a small
improvement.
We call the data after binning ``binned intensity data,''
as in this form the variables correspond to the
intensity (height) of peaks in particular $m/z$ bins.
We further reduce the ``binned intensity data'' to
``binary binned data'' where the variables ($m/z$ bins)
are binary (one/zero) valued---corresponding to
presence/absence of peaks as described by
\citet{Koch2013}, Example~6.12.
The number of $m/z$ bins in the binned data will vary
depending on the choice of bin size.
Our analyses are not sensitive to choice of bin size,
provided the bin size chosen is within a reasonable
range (0.05--2~$m/z$).
We use an intermediate bin size of $0.25$
$m/z$, which yields
$5891$ (nonempty)
$m/z$ bins in the {motivating dataset}.
The principal effect that choice of bin size has is on
the total number of $m/z$ bins, and the number that are
removed in the dimension reducing steps of our method---smaller bin sizes will result in more $m/z$ bins
initially, and more being removed, larger bin sizes
result in fewer $m/z$ bins initially, and fewer being
removed.

The use of the binary data has a number of advantages,
including separation of tissue types, computational
efficiency, and allowing the use of the easily
interpretable DIPPS approach we describe in
Section~\ref{DIPS}.
Extraneous variables such as matrix crystal morphology
can have an adverse effect on the intensity
measurements by adding noise [\citet{Garden2000}].
\citet{Deininger2011} attempt to address these effects
by normalization of their intensity values.
One additional advantage of using the binary
transformation suggested by
[\citet{Koch2013}, Example~6.12] is that it circumvents
the effects of such extraneous variables by avoiding
direct use of the intensity values, and in this sense
can be considered a data cleaning step.
All subsequent analyses will concern only the
binary data, as indicated in Figure~\ref{flowchart}, and we
will refer to these simply as ``the data'' from
here on. Similarly, we will refer to the variables in
these data as ``$m/z$ bins,'' as we have done in this
section.

\section{Clustering spatially smoothed data}
\label{clus}

We propose a two-step method for separating spectra
into groups corresponding to tissue types:
first, a smoothing step which acts as a data cleaning
and dimension reduction step, and incorporates the
spatial information from MALDI-IMS into the data.
Second, a clustering step.

\subsection{Smoothing step}
\label{smoothing}

The spatial information available in MALDI-IMS data can
be used to clean the data and remove $m/z$ bins that
are spatially dispersed.
We incorporate this spatial information through a
spatial smooth.

Let $\mData$ be a $d \times n$ binary matrix; the rows
of $\mData$ correspond to $m/z$ bins and are denoted
$\vDataRow{i}$, the columns of $\mData$ correspond to
spectra and are denoted $\vDataCol{j}$, and the entries
of $\mData$ are denoted $\sData{i}{j}$.
These entries $\sData{i}{j}$ take the value one if
peaks are present, and zero if no peaks are present in
the $m/z$ bin $i$ in spectrum $j$.
Let $0 \leq\tau< \frac{1}{2}$ be a smoothing
parameter and $\delta\geq0$ a distance cutoff.

We iteratively update the values of $\mData$.
Let $\mData^{(k)}$ denote the updated matrix at the
$k$th iteration.
Similarly, let $\vDataRow{i}^{(k)}$,
$\vDataCol{j}^{(k)}$ and $\sData{i}{j}^{(k)}$ denote
the rows, columns and values of $\mData^{(k)}$,
respectively.
At the $k$th iteration, the proportion of spectra,
$\sSmoothProp{i}{j}^{(k)}$, in a spatial
$\delta$-neighborhood of the $j$th spectrum
$\vDataCol{j}^{(k)}$ whose values at the $i$th $m/z$
bin $\vDataRow{i}^{(k)}$ agree with
$\sData{i}{j}^{(k)}$ is
%
\begin{equation}
\sSmoothProp{i} {j}^{(k)} = \biggl\{ \bigl(1 - \sData{i}
{j}^{(k-1)} \bigr) + \bigl(2\sData{i} {j}^{(k-1)} - 1 \bigr) \biggl(
\frac{\vDataRow{i}^{(k-1)} \vDistanceNeighbourhood
{j}^\intercal- \sData{i}{j}^{(k-1)}}{\mathbf{1}_{1 \times n}
\vDistanceNeighbourhood{j}^\intercal- 1} \biggr) \biggr\}. \label{eqnTij}
\end{equation}
This\vspace*{1pt} proportion determines if the value
$\sData{i}{j}^{(k)}$ should be changed.
In (\ref{eqnTij}), $\vDistanceNeighbourhood{j}$ denotes a
$1 \times n$ indicator vector with entry 1 if the
corresponding indexed spectrum is in a
$\delta$-neighborhood of $\vDataCol{j}$ and zero
otherwise.
If this proportion $\sSmoothProp{i}{j}^{(k)}$ is less
than $\tau$, we update the value $\sData{i}{j}^{(k)}$
as in (\ref{eqnXk}).

We generate the smoothed data by iteratively
calculating the entries of $\mData^{(k)}$:
%
\begin{equation}
\label{eqnXk} \sData{i} {j}^{(k)} = \cases{ \sData{i} {j}^{(k-1)},
&\quad if $\sSmoothProp{i} {j}^{(k)} > \tau $,
\cr
1 - \sData{i}
{j}^{(k-1)}, &\quad if $\sSmoothProp{i} {j}^{(k)} \leq \tau$,}
\end{equation}
for $k = 1, 2, \ldots,$ starting with
$\mData^{(0)} = \mData$.
We stop when convergence is reached, that is, when
$k = k^* = \min \{ k: \mData^{(k)} = \mData^{(k-1)}  \}$.
The spatially smoothed data are~$\mData^{(k^*)}$.

Remarks on the smoothing process:
\begin{longlist}[2.]
\item[1.] Without\vspace*{1pt} loss of generality, we let the distance
between adjacent pixels be one.
We choose $\delta= \sqrt{2}$ which results in a
range $1$ \emph{Moore neighborhood};
see \citet{Gray2003}.
This neighborhood is used in the cellular automata
literature including \citet{Conway1970}.
It is worth noting\vspace*{1pt} that acquiring the range $1$ Moore
neighborhood by using the Euclidean distance and
$\delta= \sqrt{2}$ on a regular grid is equivalent
to using the Tchebychev distance, and $\delta= 1$.

\item[2.] The smoothing parameter $\tau$ defines the
proportion of neighboring spectra needed to agree in
order for a particular value to remain unchanged at
any given step.
Small values of $\tau$ smooth less ($\tau= 0$ leaves
the data unmodified), while larger values smooth
more.
Results will not significantly change if $\tau$ is
within the same $\frac{1}{8}$-wide interval, as
changing $\tau$ within these intervals will affect
spectra only on the boundary of the acquisition
region (spectra with less than $8$ neighbors).
The limit $\tau\rightarrow\frac{1}{2}$ results in
maximum smoothing and is equivalent to the intuitive
median smooth.
In practice, the median smooth tends to yield
over-smoothed data and often fails to converge.
We chose an intermediate smoothing parameter,
$\tau= \frac{1}{4}$, for these analyses.
The values $\frac{1}{8}$ and $\frac{3}{8}$ could also
be used, for less or more smoothing, respectively.

\item[3.] Alternative smoothing options include kernel
methods [\citet{Wand1995}] which apply to continuous
data.
These methods produce continuous values when applied
to binary data, for which there is no clear
interpretation.
Our method produces binary smoothed data,
maintaining the interpretability of the binary
values.

\item[4.] At each smoothing iteration $k$, $m/z$ bins are
smoothed independently, and within each $m/z$ bin all
observations are smoothed simultaneously at each
step.
This means that it is possible to parallelize the
smoothing algorithm, making efficient use of
computational resources.

\item[5.] Our smoothing step plays a similar role in our
approach to the combined two-step method of
\citet{Alexandrov2013a}.
\citet{Alexandrov2013a} use first an edge-preserving
smooth [\citet{Tomasi1998}] and then a measure of
spatial chaos to remove spatially chaotic images.
Our method also removes spatially chaotic images by
reducing them to empty.
Improvements could potentially be achieved by
combining the two approaches.
\end{longlist}


Bins that exhibit occurrence of peaks in a small number
of spatially delocalized spectra or in almost all
spectra constitute a large proportion of all $m/z$
bins.
These bins tend not to be relevant, as they are usually
internal calibrants [\citet{Gustafsson2012}], errors,
contaminants or tissue regions that are too small to
be of interest due to the spatial (lateral) resolution
used (100~\textmu m).
This last point could be improved by using a finer
lateral resolution, as discussed by \citet{Schober2012}.
In these data, however, biological structures which are
the same size as or smaller than the acquisition
resolution, in this case 100~\textmu m, will be removed
by the smoothing.
These $m/z$ bins have zero variance after the smoothing
step.
Following the smoothing, these zero-variance $m/z$ bins
are removed, reducing the dimension of the data.
The {motivating dataset} has
$5891$ $m/z$ bins
before the smoothing step. After the smoothing step
$1022$ of
these $m/z$ bins have nonzero variance and the
remainder are discarded.

\subsection{Clustering step}
\label{clustering}

The second step in our approach concerns clustering of
the spatially smoothed data.
We use $k$-means clustering.
Based on the information available from the histology,
there are three broad tissue types present which could
be labeled as fatty, connective and cancer tissue,
respectively.
Further, there are spectra that were acquired
off-tissue.
Thus, we perform $k$-means clustering with $k=4$.
We choose initial cluster centers at random from the
sample, repeat this process $100$ times and choose the
clustering with minimum within-cluster sum of
spectra-to-centroid distances.

$k$-means clustering of the binned intensity data with
the default Euclidean distance does not lead to
interpretable or spatially localized clusters
[\citet{Koch2013}, Example 6.12].
In contrast, $k$-means clustering of the binned
intensity data with the cosine distance, and of the
binary data with the Euclidean or cosine distance, leads
to clusters that correspond to the different tissue
regions.
Since there is no clear superiority of one distance
over the other for the binary data, we continue with
only the cosine distance, which has the added advantage
of having become an established measure of closeness
for high-dimensional data and associated consistency
results [see
\citet{Koch2013}, Sections~2.7, 13.3 and 13.4, and
references therein].


For data with associated spatial information (such as
MALDI-IMS data), it is natural to display the cluster
membership in the form of cluster maps or cluster
images: colored pixels at the $(x,y)$ coordinates of
the spectra which show the cluster membership of each
spectrum using different colors to identify clusters.
The H$\&$E stained tissue cross-section and result of
$4$-means cluster analysis (by cosine distance) of the
{motivating dataset} are shown in Figure~\ref{Cos4clus}(a)~and~(b), respectively.
The cluster membership in Figure~\ref{Cos4clus}(b)
corresponds well with tissue types as determined by the
histology in Figure~\ref{Cos4clus}(a): \colboxadipose
corresponds to fatty tissue, \colboxstroma to
cancer-associated connective tissue and \colboxcancer
to four cancer nodules [indicated by arrows in
Figure~\ref{Cos4clus}(a)].
The fourth cluster in \colboxofftissue corresponds well
with off-tissue spectra, apart from a small amount of
``bleed-out'' from the \colboxadipose cluster possibly
caused by nonspecific or ``leakage'' analytes.
The correspondence between cluster results and
histology demonstrates that the spatial smooth and
cluster analysis isolate key molecular information
which allows differentiation of tissue types by their
mass spectra.

This correspondence between cluster results and
histology, particularly for the cancer tissue-type, is
important for the interpretation of results that follow,
and so we took extra steps to validate its accuracy.
A pathologist annotated the H$\&$E stained tissue
section for the {motivating dataset} [shown in
Figure~\ref{Cos4clus}(a)], indicating regions of cancerous
tissue.
In order to avoid ambiguity in annotation, we then
created an annotation subset identifying spectra whose
origin is unambiguously cancerous tissue (omitting
spectra from tissue regions of mixed tissue types, or
tissue of ambiguous type).
This allows us to be confident that spectra in this
annotation subset are definitely from cancerous tissue,
giving us a diagnostic measure of accuracy for our
cancer cluster.
The annotation subset we obtained for the
{motivating dataset} contained
$515$ spectra,
$499$
($97$\%)
of which where also contained in the
$778$ spectra of
the \colboxcancer cancer cluster of
Figure~\ref{Cos4clus}(b).

\section{The difference in proportions of occurrence (DIPPS) approach}
\label{DIPS}

In Section~\ref{clus} we mention the good agreement between
tissue types visible in the histology of
Figure~\ref{Cos4clus}(a) and clusters of
Figure~\ref{Cos4clus}(b).
From the biological perspective it is of great interest
to be able to quantify the differences between these
tissue types in an easily interpretable way.
At a mathematical level, a characterization of the
differences between the tissue types translates into an
identification of $m/z$ bins that discriminate them.
We propose the DIPPS approach for identifying
discriminating $m/z$ bins.
Other methods for determining $m/z$ values exist in the
supervised learning literature, for example, support
vector machines and PCA-based linear discriminant
analysis.
A comprehensive comparison of the DIPPS approach with
these methods in determining the ``best'' and ``correct
number'' of discriminating $m/z$ values is needed, but
such a comparison is beyond the scope of this paper.
Instead we restrict attention to the new DIPPS approach
which we first explain for an arbitrary subset of
binary data, and then apply to the {motivating dataset}
using spectra from the cancer cluster as the subset of
interest, as distinguishing these spectra from the
noncancer spectra is particularly important.

We define the DIPPS statistic for a fixed subset of
data in (\ref{eqnDIPS}), and show how this new statistic
introduces a ranking of the $m/z$ bins based on their
ability to characterize the subset of interest.
This DIPPS statistic leads to a natural heuristic,
introduced in (\ref{eqncutoff}), for selecting a number
of the ranked $m/z$ bins, which we call \emph{DIPPS
features}, that ``best'' characterize the subset of
interest.
The extraction of these DIPPS features can be
interpreted as a dimension reduction step.
For data with spatial meta-data, such as MALDI-IMS
data, we propose a way of displaying graphically the
information obtained from the statistic in an easily
interpretable summary image.
These maps make this technique useful in exploratory
analyses, as combining the DIPPS features into a
single interpretable image allows for broad conclusions
to be drawn quickly and easily.
When the amount of data becomes large, the approach
commonly used in proteomics, namely, of considering each
$m/z$ bin individually, is of limited use and the
ability to produce a single image that summarizes many
$m/z$ bins becomes particularly useful.

Let $\subsetCol$ be a subset of observations (columns)
of the data, $\mData$.
We let $\mathbf{p}(\subsetCol)$ denote the mean of the
observations in $\subsetCol$ and, similarly, let
$\mathbf{p}(\subsetCol^c)$ denote the mean of the
observations in its compliment $\subsetCol^c$.
As the data are binary, the $k$th entry of the
vector $\mathbf{p}(\subsetCol)$ is the proportion of
observations in $\subsetCol$ that take the value one
for the $k$th $m/z$ bin.
The interpretation of binary data as the occurrence of
an event---here existence of peaks---allows the mean
$\mathbf{p}(\subsetCol)$ to be interpreted as the
proportions of occurrence (for each $m/z$ bin) in
$\subsetCol$.
Considering occurrence (presence of peaks) in each
$m/z$ bin as a predictor of which spectra should be in
$\subsetCol$ allows us to interpret the corresponding
entry of $\mathbf{p}(\subsetCol)$ as the ``sensitivity'' or
true positive rate of this prediction.
Similarly, each entry of the vector
$\mathbf{1}_{d \times1} - \mathbf{p}(\subsetCol^c)$ is the
proportion of observations in $\subsetCol^c$ that take
the value zero for the corresponding $m/z$ bin, and,
from the perspective of treating each $m/z$ bin as a
predictor for which spectra should be in $\subsetCol$,
can be considered the ``specificity'' or true negative
rate.
In order to characterize $\subsetCol$, both sensitivity
and specificity should be high.
We sum these measures of sensitivity and specificity,
and subtract one to give a range of $[-1,1]$ and define
the vector of DIPPS statistics $\vDIPS$ for $\subsetCol$ as
%
\begin{equation}
\vDIPS(\subsetCol) = \mathbf{p}(\subsetCol) - \mathbf {p}\bigl(
\subsetCol^c\bigr). \label{eqnDIPS}
\end{equation}
For convenience of notation we omit the dependence on
$\subsetCol$, and write $\vDIPS(\subsetCol) = \vDIPS$.
We will similarly omit the $\subsetCol$ dependence for
$\vTemplate{\sCutoff}$, $n_\sCutoff$, $\vCentroid$ and
$a^*$ below, as we are treating
$\subsetCol$ as fixed.
We use the entries of the vector $\vDIPS$ to rank the
$m/z$ bins: the entry with the greatest value
corresponds to the $m/z$ bin that characterizes
$\subsetCol$ best.

Next we determine the set of $m/z$ bins that
collectively best characterize $\subsetCol$, that is,
the DIPPS features.
We do this by finding a cutoff value in a data-driven
way as follows.
For $a > 0$ let $\vTemplate{\sCutoff}$ be the
$d \times1$ vector that takes the value one if the
corresponding element of $\vDIPS$ is $\geq\sCutoff$,
and takes the value zero otherwise.
Let $n_\sCutoff$ be the number of entries in the vector
$\vTemplate{\sCutoff}$ equal to one.
Let $\vCentroid$ be the centroid of $\subsetCol$.
For the cosine distance, $D$, this is the average of
the normalized (to length one) vectors.
We use the cutoff
%
\begin{equation}
a^*= \argmin{\sCutoff} \bigl\{ D ( \vCentroid,\vTemplate{\sCutoff} ) \bigr\}.
\label{eqncutoff}
\end{equation}
$\vTemplate{\sCutoff}$ is a binary ``template'' vector
for representing observations in $\subsetCol$.
The centroid $\vCentroid$ represents the ``center'' of
observations in $\subsetCol$.
In (\ref{eqncutoff}) we choose $a^*$ such
that $\vTemplate{a^*}$ is as close to
$\vCentroid$ as possible.
The $n_{a^*}$ DIPPS features are the
$m/z$ bins whose corresponding entry in
$\vTemplate{a^*}$ is one.

\begin{figure}[t]

\includegraphics{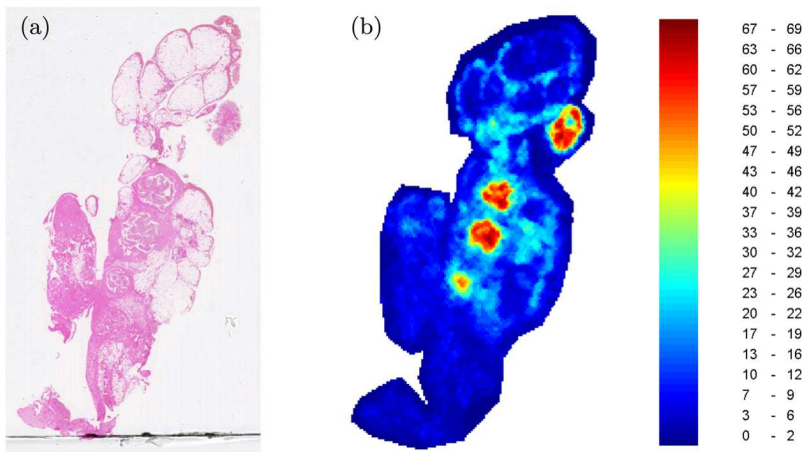}

\caption{H$\&$E stained tissue section \textup{(a)} and DIPPS map \textup{(b)} for
the cancer cluster of the motivating dataset.
The DIPPS map shows the sum of the
$n_{a^*} = 70$
$m/z$ bins with DIPPS score greater than
$a^*= 0.126$. The legend relates the counts in heat colours: cold
(blue) indicating spectra in which none of the selected
`DIPPS feature' $m/z$ bins contain peaks, hot (red)
indicating spectra in which all ``DIPPS feature'' $m/z$
bins contain peaks$\ldots.$}\label{DIPSheatmap}
\end{figure}

In the {motivating dataset} we are interested in
characterizing the cancer spectra, and thus we choose
$\subsetCol$ to be the set of spectra belonging to the
\colboxcancer cluster shown in Figure~\ref{Cos4clus}(b).
The cutoff of (\ref{eqncutoff}) is
$a^*= 0.126$
and results in
$n_{a^*} = 70$
DIPPS features being selected from the
$1022$ $m/z$ bins remaining after
smoothing.
For each spectrum
$\vDataCol{j}$, the sum of the DIPPS features
$\vTemplate{a^*}^\intercal\vDataCol{j}$,
can be visualized spatially in a ``DIPPS map,'' as shown
in Figure~\ref{DIPSheatmap}(b).
We construct the DIPPS map pointwise at each $(x,y)$
location.
The value of the DIPPS map at each $(x,y)$ location, or
pixel, represents the number of DIPPS features
exhibiting occurrence for the spectrum at $(x,y)$.
These counts are visualized in heat colours: cold
(blue) indicating spectra in which none of the selected
`DIPPS feature' $m/z$ bins contain peaks, hot (red)
indicating spectra in which all `DIPPS feature' $m/z$
bins contain peaks.
The ability to visualize results in a DIPPS map, which
is easy to interpret, is attractive, as considering so
many $m/z$ bins individually can be time consuming and
fail to provide a ``big-picture'' perspective.

%

The $70$ DIPPS
features characterize the cancer tissue, and thus
deserve inspection.
Follow-up analysis can be done to identify the peptides
that these $m/z$ bins correspond to, and to draw
inference as to their parent proteins.
To illustrate the biological significance of these
$m/z$ bins, we compared their $m/z$ values to the
previously published results of
\citet{Gustafsson2012thesis}.
Two of the highly ranked $m/z$ bins (listed in
Table~\ref{tabmassIDs}, along with inferred parent proteins)
were previously identified as highly expressed in
cancer by manual assessment of spatial distributions.
The identifications were achieved through both mass
matching to LC-MS/MS as well as {in situ} MS/MS
[\citet{Gustafsson2012thesis}].
The identity and histological distribution of these
analytes (corresponding to $m/z$ bins in our data) were
successfully validated using immunohistochemistry (see Supplement~\textup{A} [\citet{IHC}]).
This confirms that the DIPPS approach can find features
of known importance.
Crucially, the DIPPS approach produces a list of
characterizing $m/z$ bins more rapidly and
comprehensively than manual inspection of individual
$m/z$ bins.


A DIPPS map such as that shown in
Figure~\ref{DIPSheatmap}(b) has an intuitive interpretation
that the results of cluster analysis do not.
The DIPPS map highlights gradations which reveal finer
detail than is possible in cluster maps.
We discuss this point further in
Section~\ref{datasetComparison}.
This visualization using DIPPS maps becomes
increasingly important in exploratory analyses when the
number of patients and datasets increases, as it quickly
becomes infeasible to consider each of the selected
$m/z$ bins individually.
The DIPPS approach also allows $m/z$ bins crucial to
cluster/tissue differentiation to be isolated and
summarized.
This selection of DIPPS features inherent in the DIPPS
approach can also be thought of as a variable reduction
step by reducing the data to
70 $m/z$ bins.
More importantly, it successfully separates tissue type-specific $m/z$
bins, addressing the heterogeneity of
the tissue.
This facilitates the comparison of datasets, which is
the focus of Section~\ref{datasetComparison}.

\section{Application to multiple datasets}
\label{datasetComparison}

In this section we consider nine data\-sets, including
the motivating dataset.
Of these datasets, three are from each of three
different patients which we will refer to as patients
A, B and C, respectively.
We will refer to the three datasets from patient A as
A1, A2 and A3, and similarly for the datasets from
patients B and C.
Dataset A1 is the {motivating dataset}.
Each of the three datasets arising from the same
patient is acquired from thin (6~\textmu m) tissue sections
of a single surgically excised tissue.
Because of this experimental setup, we expect the
cluster and DIPPS maps of datasets from the same cancer
to be similar in terms of the location of the cancer
clusters and the selected $m/z$ bins.
We aim to separate within-patient from between-patient
variability among the datasets.

%
\begin{figure}

\includegraphics{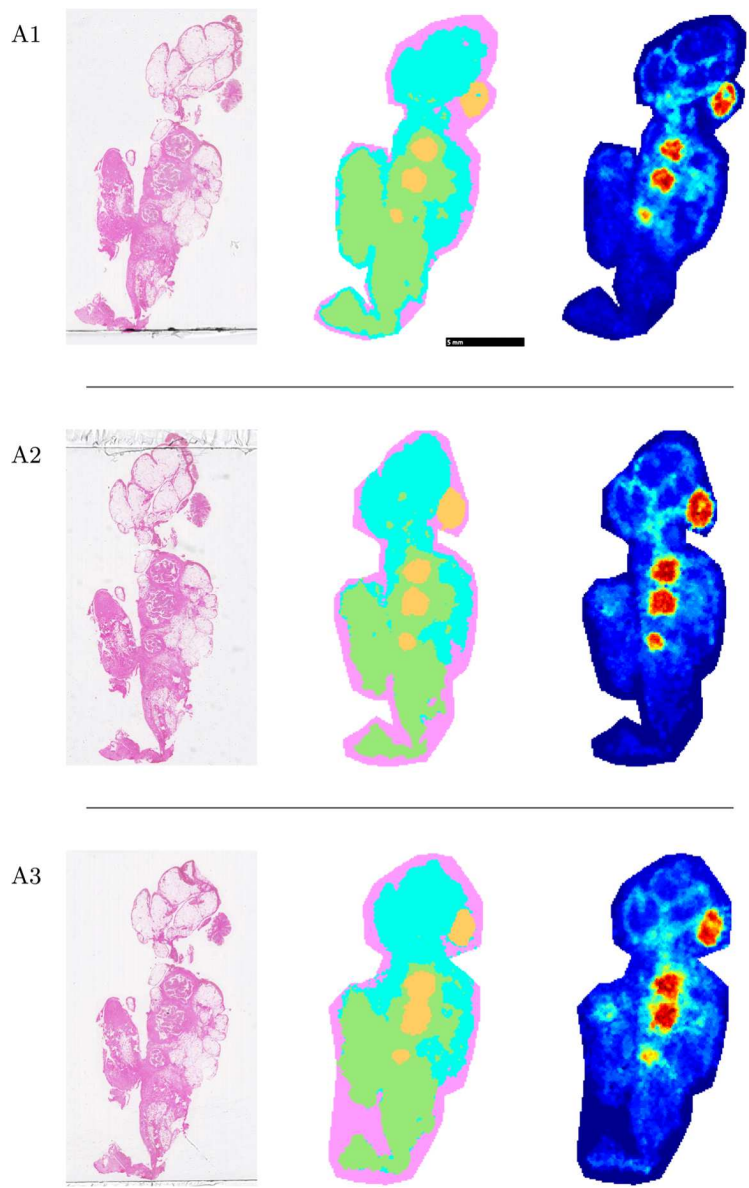}

\caption{H$\&$E stains (left column), cluster maps
(center column), and DIPPS maps for the cancer cluster
(right column) for the three datasets \textup{A1}, \textup{A2}, and \textup{A3}
(each represented in a row).
Note that $n_{a^*} = 70, 45, 61$
$m/z$ bins are visualized in the DIPPS maps for the
datasets \textup{A1}, \textup{A2}, and \textup{A3}, respectively.}
\label{additionalDatasetsA}
\end{figure}

\begin{figure}

\includegraphics{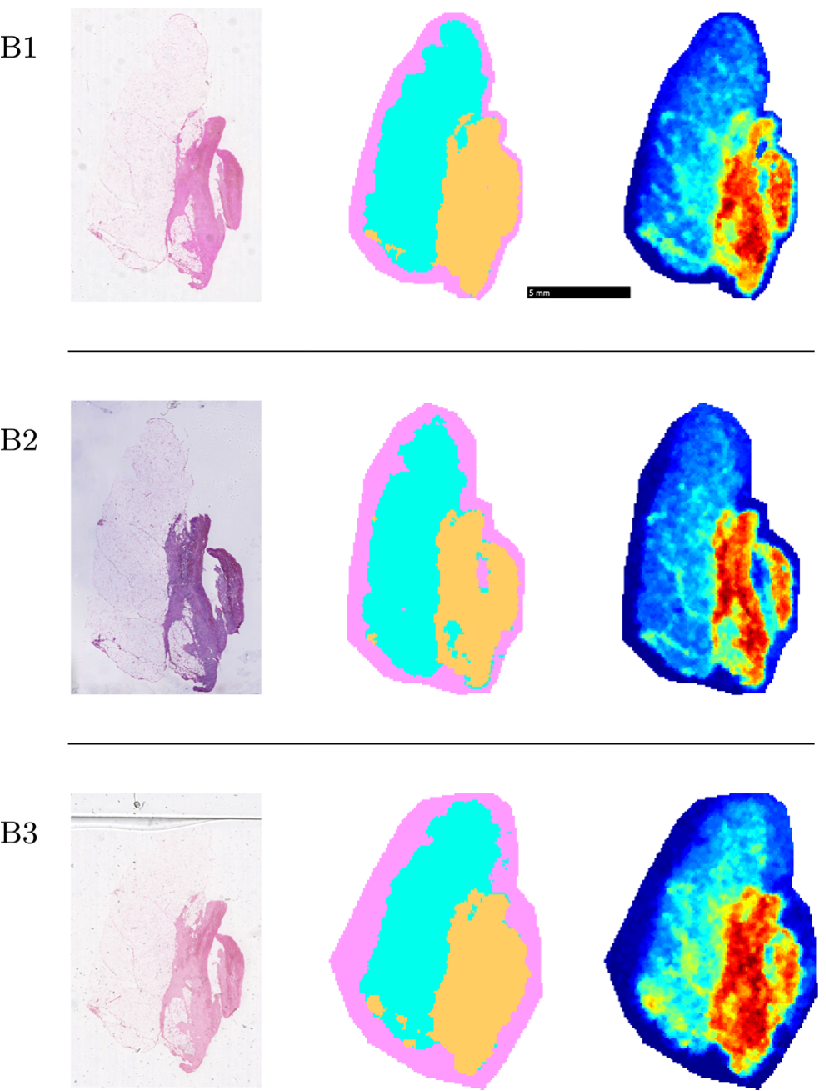}

\caption{H$\&$E stains (left column), cluster maps
(center column), and DIPPS maps for the cancer cluster
(right column) for the three datasets \textup{B1}, \textup{B2}, and \textup{B3}
(each represented in a row).
Note that
$n_{a^*} = 173, 117, 111$
$m/z$ bins are visualized in the DIPPS maps for the
datasets \textup{B1}, \textup{B2}, and \textup{B3}, respectively.}
\label{additionalDatasetsB}
\end{figure}

\begin{figure}

\includegraphics{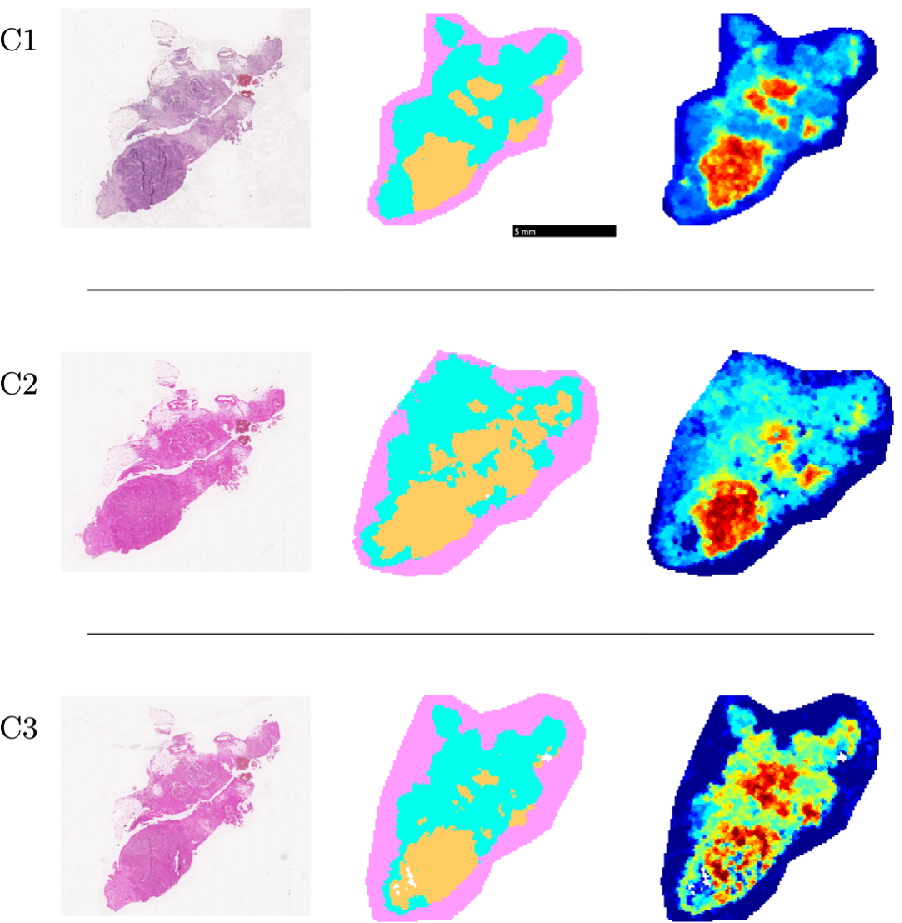}

\caption{H$\&$E stains (left column), cluster maps
(center column), and DIPPS maps for the cancer cluster
(right column) for the three datasets \textup{C1}, \textup{C2}, and \textup{C3}
(each represented in a row).
Note that
$n_{a^*} = 74, 38, 17$
$m/z$ bins are visualized in the DIPPS maps for the
datasets \textup{C1}, \textup{C2}, and \textup{C3}, respectively.}
\label{additionalDatasetsC}
\end{figure}

The results of our analyses of the nine datasets are
displayed in Figures~\ref{additionalDatasetsA},
\ref{additionalDatasetsB} and~\ref{additionalDatasetsC} for patients A, B and
C, respectively.
Each figure corresponds to one patient and shows the
three datasets in rows.
Each row consists of an H$\&$E stain on the left, a
cluster map in the center and a DIPPS map for the
cancer cluster on the right.
The first row of Figure~\ref{additionalDatasetsA} repeats
the results for the motivating dataset shown in
Figures~\ref{Cos4clus} and \ref{DIPSheatmap}.

%

In all nine datasets, as visually judged by comparison
with the H$\&$E stained images, the clustering results
correspond well with the tissue morphology.
In patients B and C the connective tissue was more
difficult to separate from the fatty tissue than in
patient A, and so $3$-means clustering was used instead
of $4$-means clustering.
Disagreements between clustering results of datasets from the
same patient serve to highlight the ability of the
DIPPS maps to find and extract information in the data
that is not available in the cluster maps, in a way
that is remarkably robust to the clustering.
For an example of this robustness property of the
DIPPS maps, consider Figure~\ref{additionalDatasetsC}---although the cluster map for dataset C2 shows a
noticeable difference in the shape of its \colboxcancer
cancer cluster, the DIPPS maps show comparatively
consistent spatial distributions. Similarly, in
Figure~\ref{additionalDatasetsB} datasets B1 and B3
show solid \colboxcancer cancer clusters, failing to
detect the clear vertical divide shown in the H$\&$E
stains, yet this divide is still apparent in the
DIPPS maps.

%

%
\begin{figure}

\includegraphics{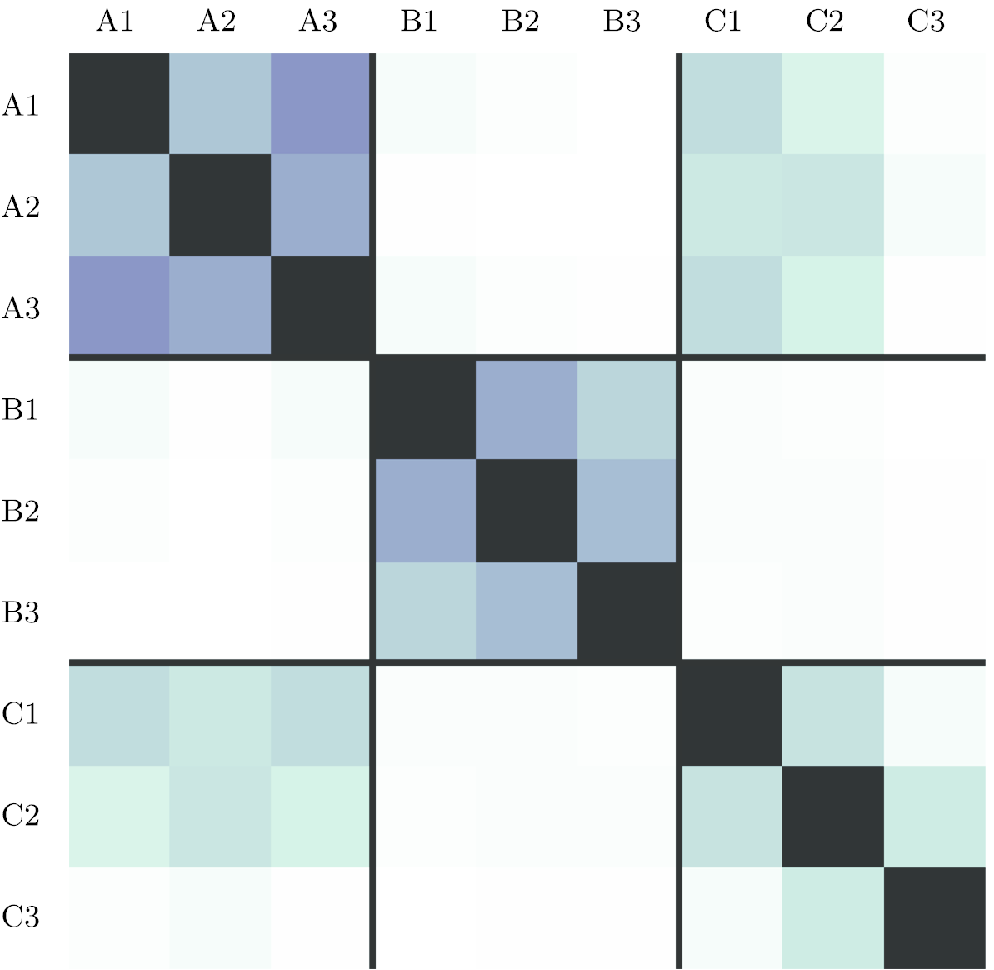}

\caption{The $9 \times9$ grid shows pairwise Jaccard
distances between the sets of $m/z$ bins characterizing
the cancer clusters.
Rows (and columns) correspond to the $9$ datasets.
The color of each pixel indicates the value of the
Jaccard distance---white corresponding to a value of
one, black to zero.}
\label{datasetComparisonfig}
\end{figure}

The DIPPS maps are also more representative of the data,
as they reveal gradual changes and fine detail that
cannot be represented by the ``hard'' boundaries in a
cluster map.
For example, consider the bottom left region of tissue
in Figure~\ref{additionalDatasetsB}.
The DIPPS maps highlight (subtly) a line following the
bottom of the tissue corresponding to heterogeneous
cancer-associated connective tissue.
This is particularly interesting, as this ``partial''
highlighting in the DIPPS map indicates that there
exists a subset of the selected $m/z$ bins that exhibit
presence in this region.




In addition to producing the DIPPS maps, the DIPPS
approach yields a set of $m/z$ bins that characterize
the subset of the data in question.
This set of DIPPS features can be used to compare
tissue types across several datasets.
We implement the DIPPS approach as described in
Section~\ref{DIPS} to identify DIPPS features for
the cancer cluster in each of the nine datasets.
To determine how similar these sets of characterizing
$m/z$ bins (DIPPS features) are, we use the Jaccard
distance [\citet{Jaccard1901}].
For a pair of sets $S_i,   S_j$, the Jaccard distance
is
%
\begin{equation}
J(S_i,S_j) = 1 - \frac{|S_i \cap S_j|}{|S_i \cup S_j|}. \label{eqnJaccard}
\end{equation}
Figure~\ref{datasetComparisonfig} shows the pairwise Jaccard
distances between the DIPPS features characterizing
cancer.
The ``colors'' range from darkest for a Jaccard distance
of zero (indicating equality of sets) to white for a
Jaccard distance of one, which indicates disjoint sets.


The block diagonal of Figure~\ref{datasetComparisonfig} is
notably dark---this illustrates that differences
within patients are smaller than differences between
patients.
This difference allows DIPPS features common to a
particular patient to be separated from features that
characterize cancer across multiple patients.
Dataset C3 is an exception to this trend, as it shows
notably less similarity to other datasets across the
board (both within patient C and to datasets from
other patients).
Further inquiry into the raw data acquisition reveals
that there was likely some problem at the sample
preparation step for this slide, possibly in either the
digestion or matrix deposition, as fewer peaks are
observed from spectra in dataset C3 in general when
compared to typical spectra from any of the other
datasets.
In this way, our proposed exploratory analyses are also
capable of alerting quality control issues, which would
not have been obvious from the clustering results
alone.

There are a number of DIPPS features common to all $3$
patients, including the $m/z$ bin centered at
$1628.75$, mentioned in Section~\ref{DIPS}.
The peptide sequence and inferred parent protein of
this feature (listed in Table~1) have been validated
by {in situ} MS/MS and immunohistochemistry (shown
in Supplement~\textup{A} [\citet{IHC}]), respectively, indicating that the
protein from which this peptide derives is highly
expressed in the cancer tissue of these patients.
This protein could therefore be investigated further as
a marker for ovarian cancer in a larger patient cohort.

Patient specific DIPPS features could be further
investigated for their ability to classify cancers
according to clinical or diagnostic criteria such as
response to treatment.
There appears to be a difference between patients A/C
and B, as can be seen in Figure~\ref{datasetComparisonfig}.
This difference could be a consequence of the
relatively larger number of DIPPS features identified
in patient B datasets.
The discrepancy in the number of selected $m/z$ bins
between patients A/C and B could be explained by the
large amount of necrotic tissue in the patient B
sample.
Mass spectra from necrotic tissue are expected to be
vastly different from those of other cancer tissue,
which would indicate that the patient-specific $m/z$
bins in patient B are likely to be markers for necrotic
tissue.

Other tissue types could be considered (fatty and
connective tissues), and a figure similar to
Figure~\ref{datasetComparisonfig} could be produced for each
of them.
Such figures do not show a notably darker
block-diagonal in the way that
Figure~\ref{datasetComparisonfig} does.
This tells us that the main differences between
patients are in the patients' cancer, rather than in
their other tissues, and reinforces how crucial it is
to address the heterogeneity of these data by
separating tissue types before conducting comparisons
between patients.

\section{Conclusion}

This paper proposes an integrated approach to
clustering and feature extraction for spatially
distributed high-dimensional data.
This approach is based on our difference in proportions
(DIPPS) statistic and includes novel visualizations
which enhance the cluster maps.
For the MALDI-IMS cancer data, these maps have a
natural interpretation in terms of the features that
characterize cancer tissue.
Application of our approach to different datasets from
a number of patients allowed us to differentiate
within-patient variability from between-patient
variability.

In proteomics, the ability to automate feature
extraction and to present these features as DIPPS maps
provides an opportunity for holistic appraisal of
MALDI-IMS data.
This is crucial due to the size and number of such
datasets and their high-dimensional nature.
By isolating features important to specific tissue
types and reporting similarities across patients, it
will be easier to identify $m/z$ bins for further
validation as tissue markers and to build models for
addressing clinical questions such as predicting
chemotherapy response and patient survival.

\section*{Acknowledgments}
The authors gratefully acknowledge the histology
annotation assistance provided by Andrew Ruszkiewicz
(SA Pathology, Adelaide, South Australia). The authors thank the
reviewers and the Editor for
helpful comments which improved the paper.



\begin{supplement}
\sname{Supplement A}
\stitle{Immunihistochemical Validation\\}
\slink[doi,text=10.1214/15-\break AOAS870SUPPA]{10.1214/15-AOAS870SUPPA} 
\sdatatype{.zip}
\sfilename{aoas870\_suppA.zip}
\sdescription{Optical images of immunohistochemical
(IHC) tissue stains, validating three proteins as
cancer-specific, including the two inferred parent
proteins of Table~\ref{tabmassIDs}.
Top row are patient A replicates, bottom row patient C replicates.}
\end{supplement}

\begin{supplement}
\sname{Supplement B}
\stitle{Source Code}
\slink[doi]{10.1214/15-AOAS870SUPPB} 
\sdatatype{.zip}
\sfilename{aoas870\_suppB.zip}
\sdescription{Sour\-ce code including cache and
intermediate data files capable of reproducing all
analyses up to and including compiling this document.
Computations where done in \texttt{MATLAB}, and results
compiled in \LaTeX{} using the \texttt{R} package \texttt{knitr}.}
\end{supplement}

\begin{supplement}
\sname{Supplement C}
\stitle{Peaklist Data}
\slink[doi]{10.1214/15-AOAS870SUPPC} 
\sdatatype{.zip}
\sfilename{aoas870\_suppC.zip}
\sdescription{Raw peaklist data, used to generate the
intermediate data files in Supplement~B [\citet{code}].}
\end{supplement}

%

\printaddresses
\end{document}